\documentclass[10pt]{article}
\pdfoutput=1
\usepackage{amsmath}
\usepackage{amssymb}
\usepackage{hyperref}
\usepackage{lmodern}  
\usepackage[utf8]{inputenx}
\usepackage[T1]{fontenc}
\usepackage{graphicx}
\usepackage{epstopdf}
\usepackage{cite}
\usepackage{color}
\usepackage[labelfont=bf,labelsep=period,justification=raggedright]{caption}
\makeatletter
\renewcommand{\@biblabel}[1]{\quad#1.}
\makeatother

\date{}
\usepackage{pdfpages}

\begin{document}
\begin{flushleft}

{\Large \textbf{Tracing the Attention of Moving Citizens} }
\\
Lingfei Wu$^{1}$, Cheng-Jun Wang$^{2}$
\\
\bf{1} Knowledge Lab, Computation Institute, University of Chicago, Chicago, IL., 60607, United States
\\
\bf{2} Computational Communication Collaboratory, School of Journalism and Communication, Nanjing University, Nanjing, 210093,  P.R. China
\\
\end{flushleft}

\section*{Abstract}
With the widespread use of mobile computing devices in contemporary society, our trajectories in the physical space and virtual world are increasingly closely connected. Using the anonymous smartphone data of $1 \times 10^5$ users in 30 days, we construct the mobility network and the attention network to study the correlations between online and offline human behaviors. In the mobility network, nodes are physical locations and edges represent the movements between locations, and in the attention network, nodes are websites and edges represent the switch of users between websites. We apply the box-covering method to renormalize the networks. The investigated network properties include the size of box $l_B$ and the number of boxes $N(l_B)$.  We find two universal classes of behaviors: the mobility network is featured by a small-world property, $N(l_B) \simeq e^{-l_B}$, whereas the attention network is characterized by a self-similar property, $N(l_B) \simeq l_B^{-\gamma}$. In particular, with the increasing of the length of box $l_B$, the degree correlation of the network changes from positive to negative which indicates that there are two layers of structure in the mobility network. We use the results of network renormalisation to detect the community and map the structure of the mobility network. Further, we locate the most relevant websites visited in these communities, and identify three typical location-based behaviors, including the shopping, dating, and taxi-calling. Finally, we offer a revised geometric network model to explain our findings in the perspective of spatial-constrained attachment. 

\section*{Introduction}
Digital media, especially Internet and smartphones, provides a new lens to study human behaviors in both the physical space and the virtual world. Researchers are getting more interested in studying the relationship between online and offline human behaviors \cite{ginsberg2009detecting, bond201261, zhao2014scaling}. For example, Ginsberg et al. tried to predict the influenza epidemics using search engine query data \cite{ginsberg2009detecting}; Bond et al. studied the relationship between Facebook use and offline political mobilization \cite{bond201261}; Zhao et al. investigated the relationship of human movements in cyberspace and physical space using a mobile phone dataset \cite{zhao2014scaling}. In particular,  Zhao et al.  found a superlinear scaling relation between the mean frequency of visit $<f>$ and its fluctuation $\sigma$, and there was a strong correlation for the average frequency of $<f>$ in both spaces \cite{zhao2014scaling}. However, the high-resolution spatial behaviors in these research remained unexplored. In this study, we use the fine-grained smartphone data of $1 \times 10^5$ users in a major city of China to study the interrelationship between online human behaviors and offline mobility in the perspective of network structures, especially the small world property and the fractal character. 

In previous research, most complex networks are found to be small-world \cite{milgram1967small, watts1998collective, albert1999internet,  latora2001efficient,  montoya2002small, bassett2006small}. However, a series of recent studies discover that many complex networks are featured by fractal patterns \cite{song2005self, goh2006skeleton, song2006origins, radicchi2008complex}. Small-world and fractality are two fundamental structural properties of complex networks. Yet, suggested by \cite{song2005self, song2006origins,rozenfeld2010small}, there is a seemingly contradictory relationship between the small-world property and the fractality property in complex networks. The small-world property of complex networks suggests that the average diameter of the network $l$ increases logarithmically with the total number of nodes $N$, that is $N \simeq e^{l/l_0}$, where $l_0$ is a characteristic length. However, the self-similarity of fractality requires a power relation between $N$ and $l$, that is $N \simeq l^{-\gamma}$. 

We construct the mobility network and the attention network to compare their network structures. In the mobility network, nodes are physical locations and edges represent the movements between locations \cite{banavar1999size, song2010limits}, and in the attention network, nodes are websites and edges represent the switch of users between websites \cite{wang2016scaling}. We use the network renormalisation method in general and the box-covering method in specific to investigate the network structure. The box-covering method is proposed by Song et al. \cite{song2005self}. It is originated from the box-counting method which is widely used in calculating the fractal dimension \cite{bunde2012fractals}. Considering a network embedded in euclidean space, we cover the network with boxes of size $l_B$, and the number of boxes needed is denoted as $N_B$. Thus the fractal dimension or box dimension $d_B$ can be given from $N_B \simeq l_B^{-d_B} $. 

Our findings reveal that the attention network is self-similar, whereas the mobility network shows a small world property. The intriguing implication is that there are structural differences between the physical mobility and online surfing for smartphone users. That is to say, human behaviors in physical space and cyberspace represent two universal classes. Surprisingly, we find that the degree correlation of the mobility network shifts from positive to negative with the increasing of the length of box $l_B$, which is found for the first time to our best knowledge, and we define and further explain such phenomenon as “spatial-constrained hub”.  

With these properties found above, we apply the community detection technique on the mobility network. To be specific, using renormalisation techniques, we map the spatial space into different communities, and measure the association between spatial communities and the websites visited by the moving citizens by computing the TF-IDF values. Thus we can identify the websites frequently visited by the users of a given community, and conversely, we can tell the communities frequently visited by the users of a given website.  Using this method, we find three examples of  location-based behaviors, including the shopping, dating, and taxi-calling. The investigations in the local structure of the mobility network also help us further understand the spatial-constrained hub. Finally, we explain the observed empirical patterns with a revised geometric network model.

  \begin{figure*}[!ht]
    \centering
    \includegraphics[scale=0.1]{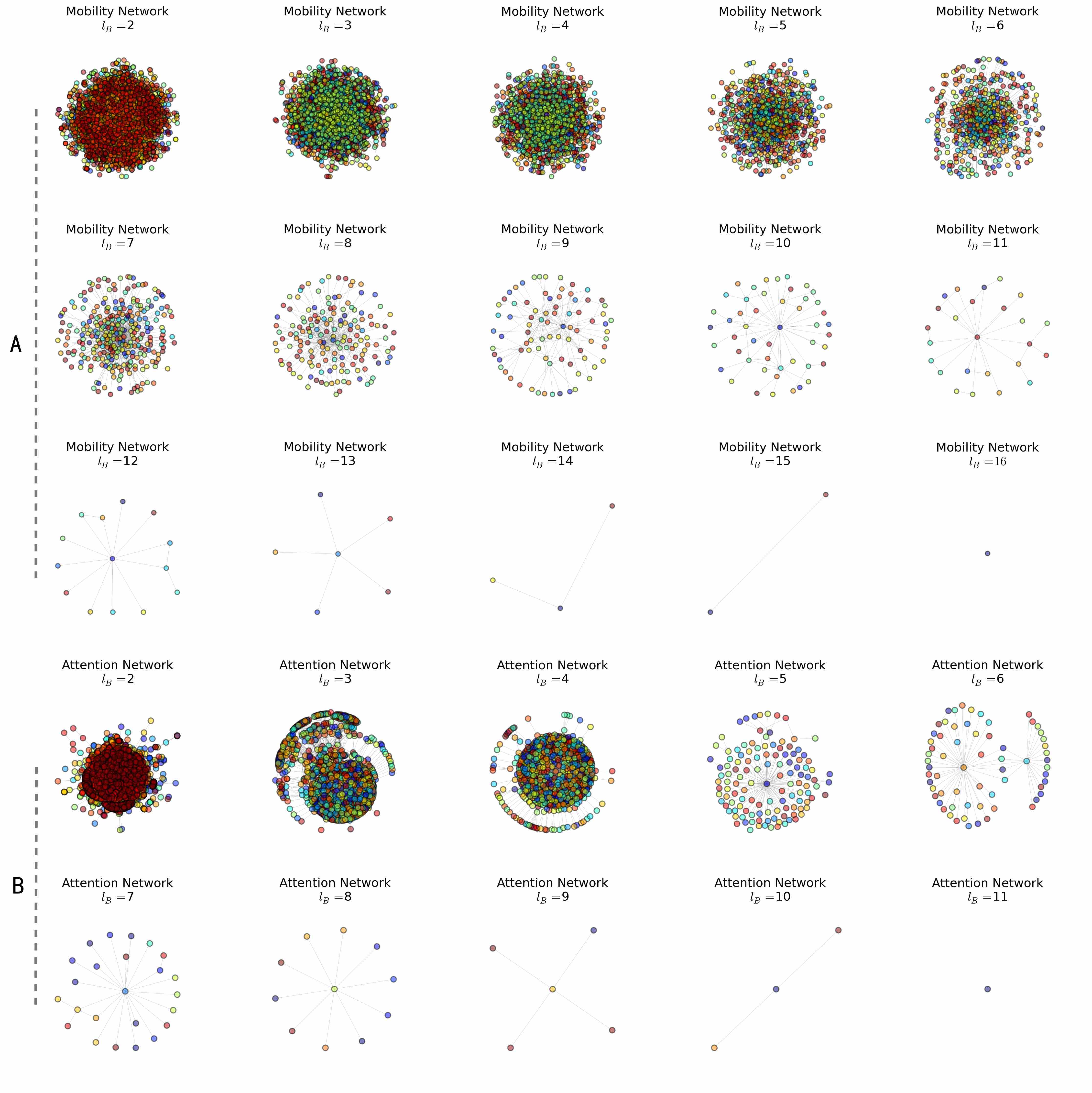}
    \caption{Renormalizations of the mobility network and attention network. A. The renormalization process of the mobility network.  It takes 15 steps to tile the mobility network into one single node.  B. The renormalization process of the attention network. It takes 10 steps to collapse the attention network into one node.}
    \label{renormalization}
  \end{figure*}

\section*{Two Universal Classes of Behaviors}
Using the box-counting method, Song et al. confirmed the self-similarity property of complex networks \cite{song2005self}. We apply the same renormalization technique in mobility networks and attention networks to investigate the fractal pattern of the network structure. Figure \ref{renormalization} presents the renormalization process for the two networks. The attention network is much more dense than the mobility network. There are 9899 nodes and 39,083 edges in the mobility network ($Density = 7.9 \times 10^{-4}$), and there are 16,476 nodes and 144,909 edges in the attention network ($Density = 10.6 \times 10^{-4}$). The diameter of the mobility network is 15, and the diameter of the attention network is only 10. Thus, it is natural that we need more boxes of the same length $l_B$ to cover the mobility network than the attention network. Actually, it takes 15 steps to tile the mobility network into a single node, and it needs 10 steps only to tile the attention network into a single node. 

  \begin{figure*}[!ht]
    \centering
    \includegraphics[scale=0.15]{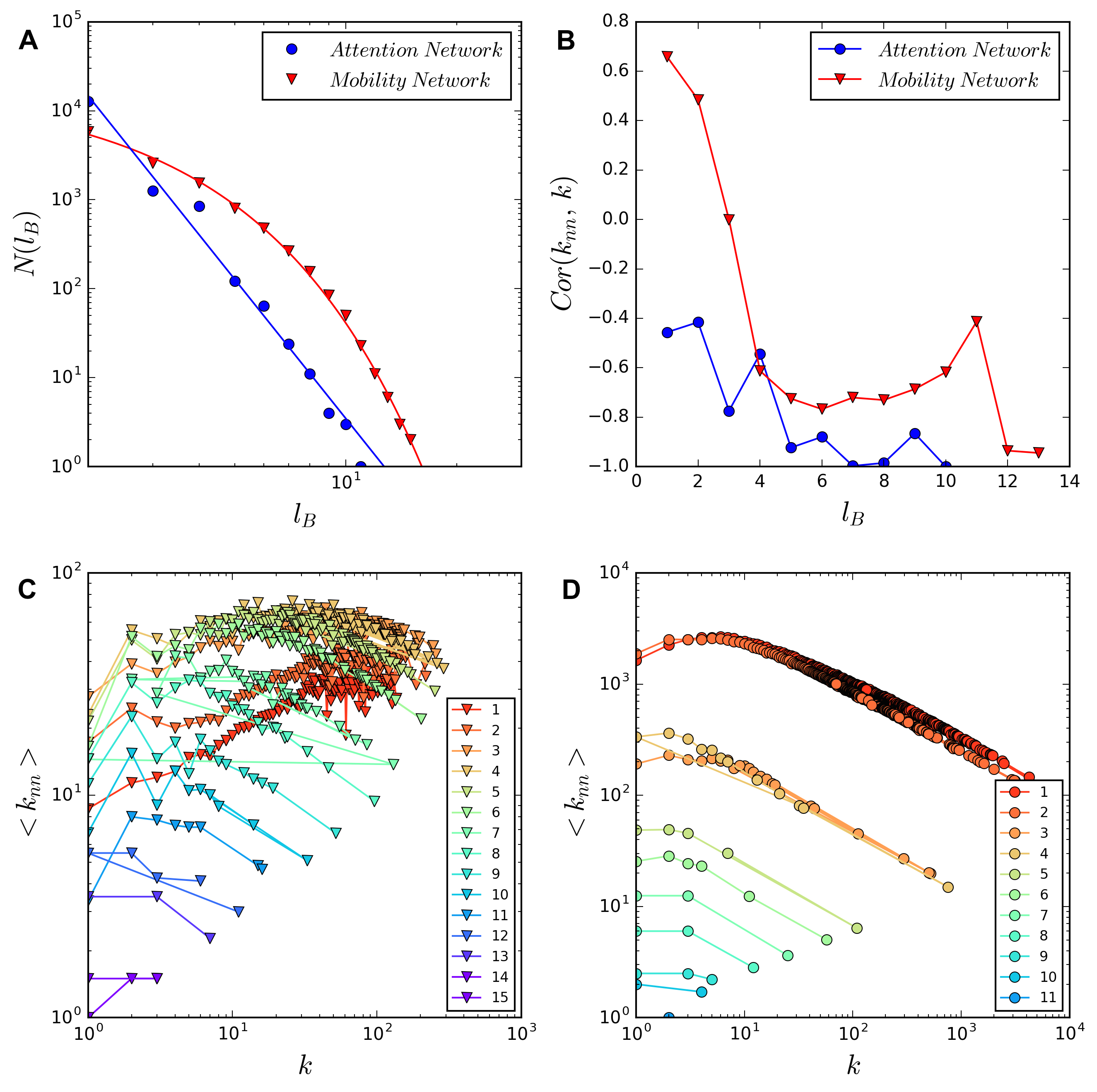}
    \caption{Two Universal Classes of Behaviors and the Transition of Degree Correlations. A. The relationship between the length of box $l_B$ and the number of boxes $N(l_B)$ reveals two universal classes of dynamics. The number of boxes of attention network $N(l_B)$ can be well fitted with a power law function of $l_B$(blue line), while the number of boxes of mobility network $N(l_B)$ can be better fit with an exponential function of $l_B$(red line). B. Degree correlation decreases when $l_B$ gets larger. When $l_B = 4$, the degree correlation gets stable. There is a transition of assortativity for the mobility networks. When $l_B$ is smaller than 3, there is a positive degree correlation $Cor(k_{nn}, k)$; When $l_B$ is larger than 2,  $Cor(k_{nn}, k)$ becomes negative. We further describe the transition process in panel C and panel D. C. For the mobility network, when $l_B$ is smaller than 3, $<k_{nn}>$ and $k$ has a positive relationship. Especially, when $l_B$ becomes larger, the positive relationship turns negative. The inset of panel C shows the change of $Cor(k_{nn}, k)$ with the increasing of $l_B$. D. For the attention network, the negative degree correlation between $<k_{nn}>$ and $k$ holds no matter how $l_B$ changes. }
    \label{transition}
  \end{figure*}

We find two universal classes of behaviors (Figure \ref{transition}B). For the attention network, the relationship between the number of boxes $N(l_B)$ and the length of box $l_B$ shows a power law function, $N(l_B) \simeq l_B^{-\gamma}$. Thus, the structure of the attention network is self-similar or fractal. Whereas for the mobility network, the number of boxes $N(l_B)$ and the length of box $l_B$ have an exponential relationship. Thus the structure of the mobility network demonstrates a small-world property. The small-world network is featured for its short diameter and a large clustering coefficient, which is efficient for transportation \cite{latora2001efficient}. This finding implies that there are two different dynamical mechanisms governing the evolution of the two networks.

\section*{Transition of Degree Correlation}

Suggested by Song et al.\cite{song2005self}, the fractal property of networks originates from disassortativity, i.e., the hubs tend not to connect directly with each other, on the contrary, the hubs tend to connect directly with non-hub nodes. Thus, the degree correlation is negative for fractal networks. To gain insights into the mechanism that leads to the two universal classes of behaviors, we further inspect the change of degree correlation against the length of box $l_B$.  The degree correlation of a network is quantified by calculating the correlation $Cor(k_{nn}, k)$ between the degree $k$ for nodes  and the average degree of their neighbors $<k_{nn}>$ \cite{maslov2002specificity}. As it has been shown in Figure \ref{transition}B, $l_B$ = 3 is a turning point at which assortativity transfers to disassortativity ($Cor(k_{nn}, k) < 0 $). The curve of $Cor(k_{nn}, k)$ for both mobility network and attention network became largely flat when $l_B$ is larger than 4. 

To better understand the transition of degree correlation under renormalization, the relationship between $<k>$ and $<k_{nn}>$ are visualized for different $l_B$. Figure \ref{transition}C-D show the detailed transition process for the mobility network and the attention network, respectively. There is a common pattern in the renormalization process for both the mobility network and the attention network.  With the increasing of $l_B$, the degree correlation $Cor(k_{nn}, k)$ and $<k_{nn}>$ decreases gradually. For example, in the attention network, when $l_B=2$, the $<k_{nn}>$ for $k=1$ is larger than 1000, while when $l_B = 6$, the $<k_{nn}>$ for $k=1$ is only 10. Figure \ref{transition}D demonstrates that the overall fractal structure of the attention network holds for different resolutions of $l_B$. On the contrary, the structure of the mobility network is more complex. There are two layers of structure in the mobility network. To tile the network with a small $l_B$ will combine the locally connected nodes into super-nodes, when $l_B$ is smaller than 3, even though the nodes within the range of 2 are collapsed into supernodes, the overall positive degree correlation (small-world property) are well maintained. When $l_B$ is larger than or equal to 3, some of these super-nodes further collapsed into much larger ones, and we call them “spatial-constrained hub”. When the spatial-constrained hub appears, the topological structure of the mobility network significantly changes from assortativity to disassortativity. According  to previous research \cite{song2005self, song2006origins}, there are smaller nodes standing between hubs. 

\section*{Location-Based Human Behaviors}

\begin{figure*}[!ht]
    \centering
    \includegraphics[scale=0.1]{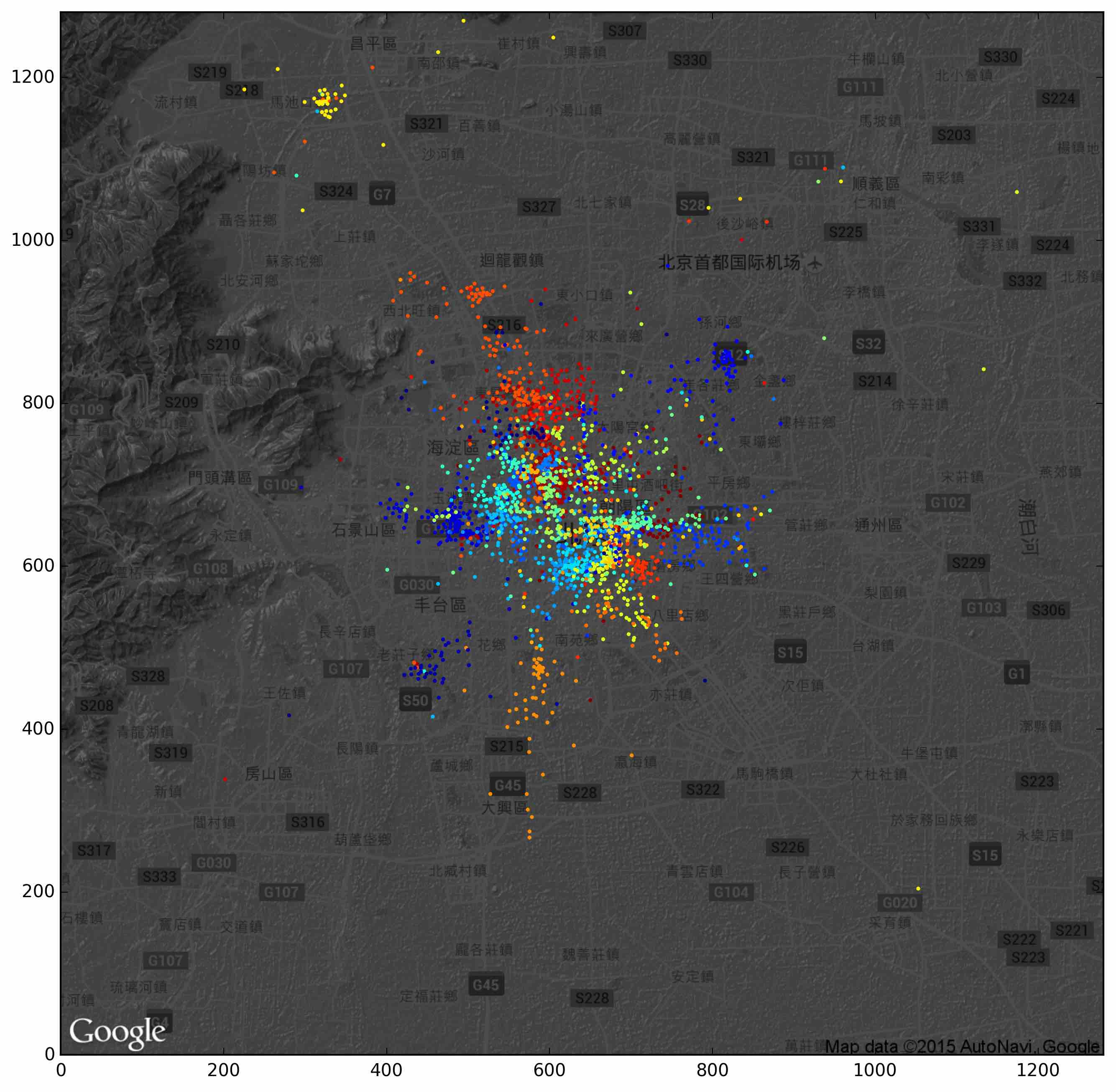}
    \caption{Community Detection of the Mobility Network. The top 20 largest communities are visualised with different colors.}
    \label{community}
  \end{figure*}

Since users are simultaneously travel in the attention network and mobility network, it is natural to infer that our online surfing behaviors within the attention network are also location-based. We will first find the communities of the mobility network, and then explore the underlying connection between the spatial communities and online surfing behaviors, in order to find out what websites users are visiting when they are moving within a community of the mobility network. Using the boxing-covering method, we can detect the communities of the mobility network. According to Figure \ref{transition}B, when $l_B = 4$, the degree correlation curve becomes relatively stable (around 0.7). Thus, we tile the mobility network into communities using the box of length 4 (see Figure \ref{community}). 

\begin{figure*}[!ht]
    \centering
    \includegraphics[scale=0.08]{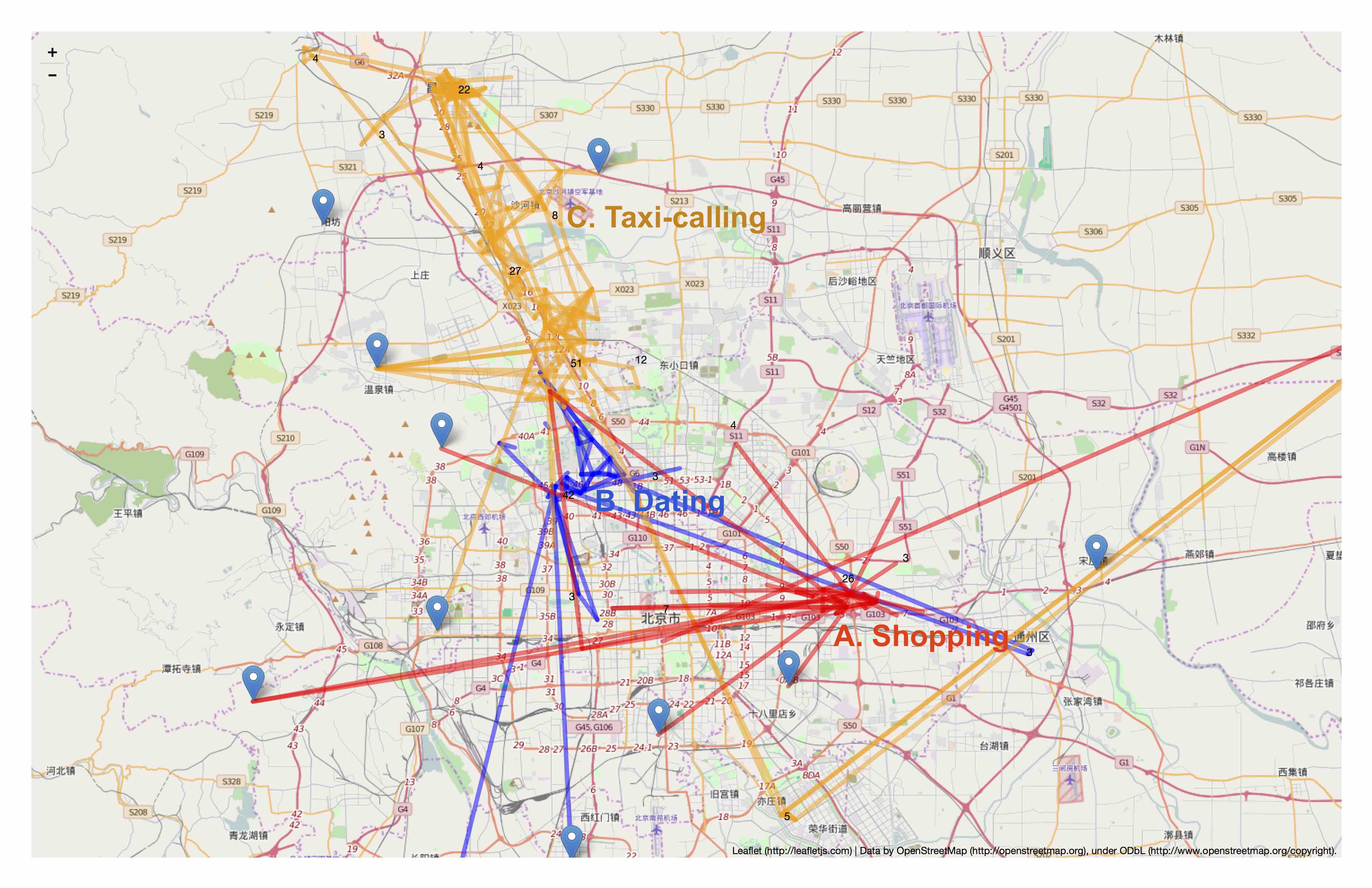}
    \caption{Geographical Distribution of Three Kinds of Mobile Internet Use Behaviors. A. Shopping (red), B. Dating (blue), and C. Taxi-calling (orange).}
    \label{three_examples}
  \end{figure*}

For each community, we calculate its TF-IDF value for all the websites. Thus, we can identify the most relevant websites for a given community, and qualitatively check the validity of the results. The assumption underlying these patterns is that the mobility behaviors in the physical space and the attention switching behaviors in the virtual space are inter-connected. As it has been demonstrated in Figure \ref{three_examples}, we find three examples of typical location-based behaviors: A. The red nodes and edges profile a community featured by the online shopping behaviors. These users move around Chaoyang Road which connects Wangfujing business street and the eastern 5th ring road. This area is around the largest business circles in Beijing and attracts a lot of consumers. With the aid of mobile internet, users can surf online to collect online information before they purchase offline. B. The blue nodes and edges mark the community which is featured by the online dating behaviors. It well depicts the footprints of online daters in Haidian district, which is the home of many well-known universities (e.g., Peking University, Tsinghua University, and Renmin University of China). C. The orange nodes and edges capture the locations of the users who use mobile texi-calling apps, especially the Didi dache app (www.xiaojukeji.com/) which is the most popular taxi apps in China. Our findings reveal that the heavy use of taxi-calling service is primarily along the highway to Changping District in the north-west of Beijing where is not well covered by the city subway.  

\section*{Geometric Network Models}
One important feature of the mobility network lies in the fact that both the node location and edge-generating mechanisms are spatial-constrained. This pattern has been well demonstrated in Figure \ref{three_examples}. We can qualitatively observe that the probability of tie-generating is constrained by the spatial distance: although long distance connections are possible, there tends to be more edges within a short physical distance, and thus we can find apparent local clustering patterns. This feature has been theorized as the spatial-constrained attachment (SCA)\cite{geometric2015}.  Based on the SCA mechanism, Zhang et. al proposed a growing geometric graph model to explain the origin of scaling behaviors in cities \cite{geometric2015}. 

In the original geometric network model, there is a geometric graph growing in d-dimensional Euclidean space. Initially, there is only one node as the seed of the growing graph located in the center of the space. At each time $t$ a new node $P$ is generated, and $P$ will be added to the graph if only it is within the radius $r$ of at least an existing node $Q$, or else $P$ will not be added to the graph. If $P$ are within the radius of more than one node $Qs$, an edge will be established between $P$ and each node of $Qs$. Since the new node $P$ will be connected with all the nodes within a given radius, this original model is named as Model all. Based on Model all, more practical extensions can be implemented. For example, to add a parameter to tune the local clustering density, or to add multiple seeds in the initial stage \cite{geometric2015}. 

 \begin{figure*}[!ht]
    \centering
    \includegraphics[scale=0.15]{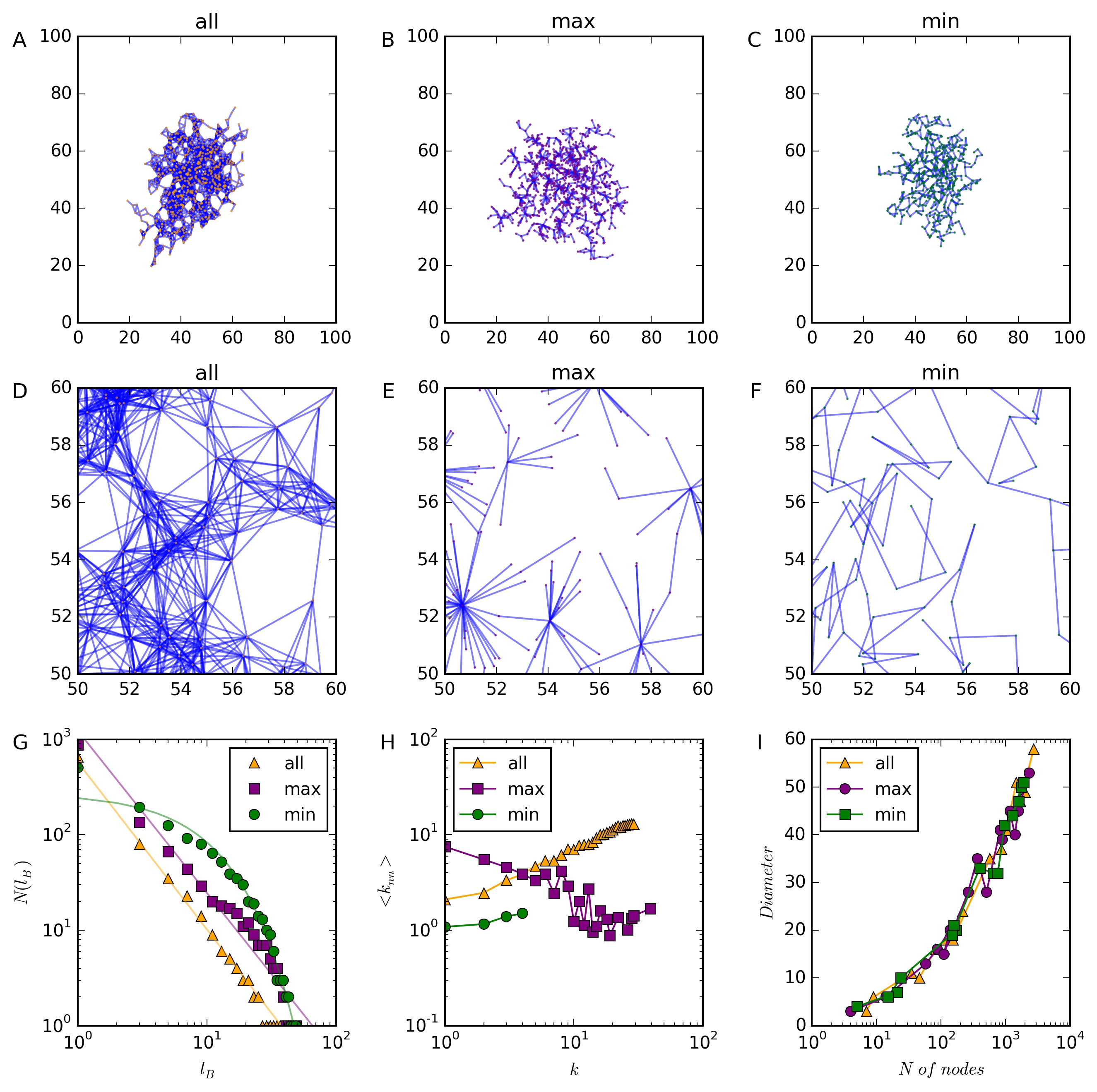}
    \caption{Geometric Network Models of Spatial-constrained Attachment. The edges between the new added node and old nodes within the radius $r$ can be established following three kinds of rules: A. the new node is connected with all the nodes within the radius (Model all). B. the new node are only connected to the node of the largest degree within the radius (Model max). C. the new node are only connected with the node of the smallest degree within the radius (Model min). Panel D, E, and F show the local network structure for the three models respectively.  G. two universal classes of behaviors are found for three kinds of geometric network models. The networks generated using Model all and Model max are fractal (the relationship between $N(l_B)$  and $l_B$ can be fitted with a power law function), and the the network generated with model min is a small-world (the relationship between $N(l_B)$  and $l_B$ can be fitted with an exponential function). H. the degree correlation of the networks generated with model all and model min is positive, while the degree correlation of the network generated with model max is negative. I. The diameter of the networks generated with three models grows exponentially with the number of nodes. }
    \label{geometric_model}
 \end{figure*}

We further extend the geometric network model by changing the edge-generating mechanism based on Model all. In Model all, a new node $P$ will be connected with all the nodes within the length of radius $r$, this rule of edge-building could be not practical in real life. One competing rule comes from the preferential attachment of the BA model \cite{barabasi1999emergence} in which new added node links preferentially to hubs. In our revised geometric models, we choose to connect the new node $P$ with only one node within the radius $r$. There are primarily two scenarios: I. Model max, to link the new node $P$ with the node $Q_{max}$ which has the largest degree within the radius $r$; II. Model min, to link the new node $P$ with the node $Q_{min}$ which has the smallest degree within the radius $r$. 

We compare the structural patterns of these three geometric network models discussed above. The panel A, B, and C of Figure \ref{geometric_model} show the overall structure of three geometric networks generated with Model all, Model max, and Model min respectively. To look into the local structure can help us gain more insights. The panel D, E, and F of Figure \ref{geometric_model} demonstrate the local connectivity generated by Model all, Model max, and Model min respectively. It is easy to find that: The geometric network generated by Model all is very dense relative to Model max and Model min (Panel D); There are significantly more hubs in the network generated with Model max, and note that there are also obvious disassortativity in the local structure (there are nodes standing between hubs, which is a feature of fractal networks, see Panel E); There are few hubs in the local structure of the network generated with Model min, and the links are somewhat like the traces of a random walk (Panel F).

We also renormalize the geometric networks generated with three models using the box-covering method. Figure \ref{geometric_model}G confirms two universal classes of behaviors by inspecting the relationship between the size of box $l_B$ and the number of boxes $N(l_B)$. The networks generated with Model all and Model min are fractal, since the relationship between $N(l_B)$ and $l_B$ can be fitted with a straight line. While the network generated with Model min shows an apparent small-world pattern that we have also found in the mobility network. We further explore the degree correlations of the generated networks. As Figure \ref{geometric_model}H has shown, there is a positive relationship between $<k_{nn}>$ and $k$ for Model all and Model min, just as what we have discovered for the mobility network. While there exists a negative relationship between $<k_{nn}>$ and $k$ for Model max. In addition, the diameters of the networks generated with three models grows identically with the number of nodes (Figure \ref{geometric_model}I). The findings confirms that model max can replicate the findings we have found in the attention network, while the model min can help explain the results of the mobility network.

\section*{Conclusion and Discussion}
To compare how human move in the virtual world and physical world has been a major interest of scientific research. Our study provides one approach to fulfill this task by constructing and comparing the mobility network and the attention network. Using networks as a mathematical structure of the complex human behaviors, we can implement the tools of network science to lend us a new lens and gain deeper insights. By renormalizing the two networks, we confirm that the attention network is featured by its fractal patterns, and the mobility network is characterized by its small-world property. The study of Zhao et al. concluded that there is a strong correlation between between human movements in cyberspace and physical world  in the perspective of flux-fluctuation law\cite{zhao2014scaling}. However, our study showed that the the human movements in two spaces represent two universal classes of behaviors. Analyzing the change of degree correlation with the size of box $l_B$ reveals that there is a surprising transition of degree correlation from positive to negative for the mobility network, while the degree correlation of the attention network stays negative with the increasing of $l_B$. This finding underlines the structural difference between online and offline human behaviors. 

Previous studies on fractal networks construct a good theoretical framework on explaining the scaling property of self-similarity. For example, Gallos et al. proposed a scaling theory of degree correlation for scale-free networks\cite{gallos2008scaling}. The degree correlation can be captured by the probability of $P(k_1, k_2)$ that two nodes of degree $k_1$ and $k_2$ are connected together by a link. Gallos et al. showed that $P(k_1, k_2)$ can be expressed as $P(k_1, k_2) \sim k_1^{-(\gamma-1)} k_2^{-\epsilon} \;\; (k_1 > k_2)$. To better estimate $\epsilon$, they proposed a quantity of $E_b(k)$ which measures to which extent the nodes of degree $k$ tend to link with more connected nodes, $E_b(k)  \equiv \frac{\int_{bk}^{\infty} P(k | k') dk'}{ \int_{bk}^{\infty} P( k') dk' }$. For the scale-free network, $E_b(k) \sim k^{-(\epsilon - \gamma)}$. By estimating the correlation exponent $\epsilon$ and the power-law exponent $\gamma$, they constructed a phase diagram which classifies scale-free networks into three areas: Below the line of $\epsilon_{random} = \gamma -1$ are the random networks (area I); When $\gamma -1 < \epsilon < 2$ is the area II where the network has a weaker hub-hub correlation than the random networks but is not fractal; Above the line of $\epsilon = 2$ is the area III where the networks are fractal. With the theory, we calculate the power exponents for the mobility network($\gamma_{mobility} = 2.66$ ) and the attention network ($\gamma_{attention} = 2.00 $), and the slopes of $E_b(k) $ for the mobility network($slope_{mobility} = 0.9$ ) and the attention network ($slope_{attention} = -0.8 $). Thus, the $\epsilon$ of the mobility network and the attention network is 1.76 and 2.80. Therefore, the mobility network is nonfractal (although its hub-hub correlation is weaker than random networks), and the attention network is fractal. However, it is also necessary to note that since the attention network and the mobility network are not perfectly scale-free network networks, such analysis may be problematic.

The transition of degree correlation uncovers the fact that the mobility network is spatial-constrained. The transition happens during the renormalization process within which the nodes covered in a box are tiled into one supernode. There are two layers of structures for the mobility network. In the initial stage of network renormalizaiton, the local clustering are not fully destroyed, thus we can still observe positive degree correlation for the mobility network (when $l_B \leqslant 2$). When $l_B > 2$, the local structure of the mobility network significantly changes from assortativity to disassortativity. Using the box-covering algorithm, the city can be classified into different communities. However, neither the original fractal model nor the classical small-world model could demonstrate the transition of degree correlation. We then inspect the geographical distribution of three location-based behaviors. The spatial-constrained feature of the edge again indicates that the mobility network is different from the attention network for its spatial-constrained mechanisms, and we need to adopt the geometric network model to explain our findings. Based on the original geometric network model (Model all), we propose two revised models (Model max and Model min) by changing the edge-generating mechanisms. The revised geometric models replicate what we have discovered in the data: both two universal classes of behaviors and the transition of degree correlation. 

Our findings about how human move in both cyberspace and physical space can help develop commercial applications to find the best spot to place online and offline geomarketing advertisement. The increasing power of search engines has made advertisement more and more precise. Computational advertising systems like Google AdSense collect and filter the attention of the most relevant users and sale it to companies. For mobile Internet use, the telecommunications operators have the most detailed information about the users’ geographical locations and the websites they are visiting. For a city, it can be separated into functional areas using the network renormalization method. Relating the moving citizens’ behaviors in cyberspace with their mobility in the physical space can help develop more accurate recommender. 

In summary, this research confirms that the human behavior in the virtual world and that in the physical world are structurally different. The human moving behaviors in the physical space and the virtual world represent two universal classes of human behaviors. There are two layers of structure in the mobility network, and only when the local structure is tiled using the box-covering method we can get a fractal structure. Connecting new nodes with non-hub nodes can help make up the two-layers structure of the mobility network. Only when the locally layer of clustered nodes are titled into spatial-constrained hubs under renormalization, the negative degree correlation can be observed.  The spatial-constrained attachment mechanism determines its unique network structure. Using the revised geometric models, we are able to replicate what we have found, and the growing of attention network and mobility network is not fully preferential attachment or anti-preferential attachment, but also spatial-constrained. 

 \begin{figure*}[!ht]
    \centering
    \includegraphics[scale=0.06]{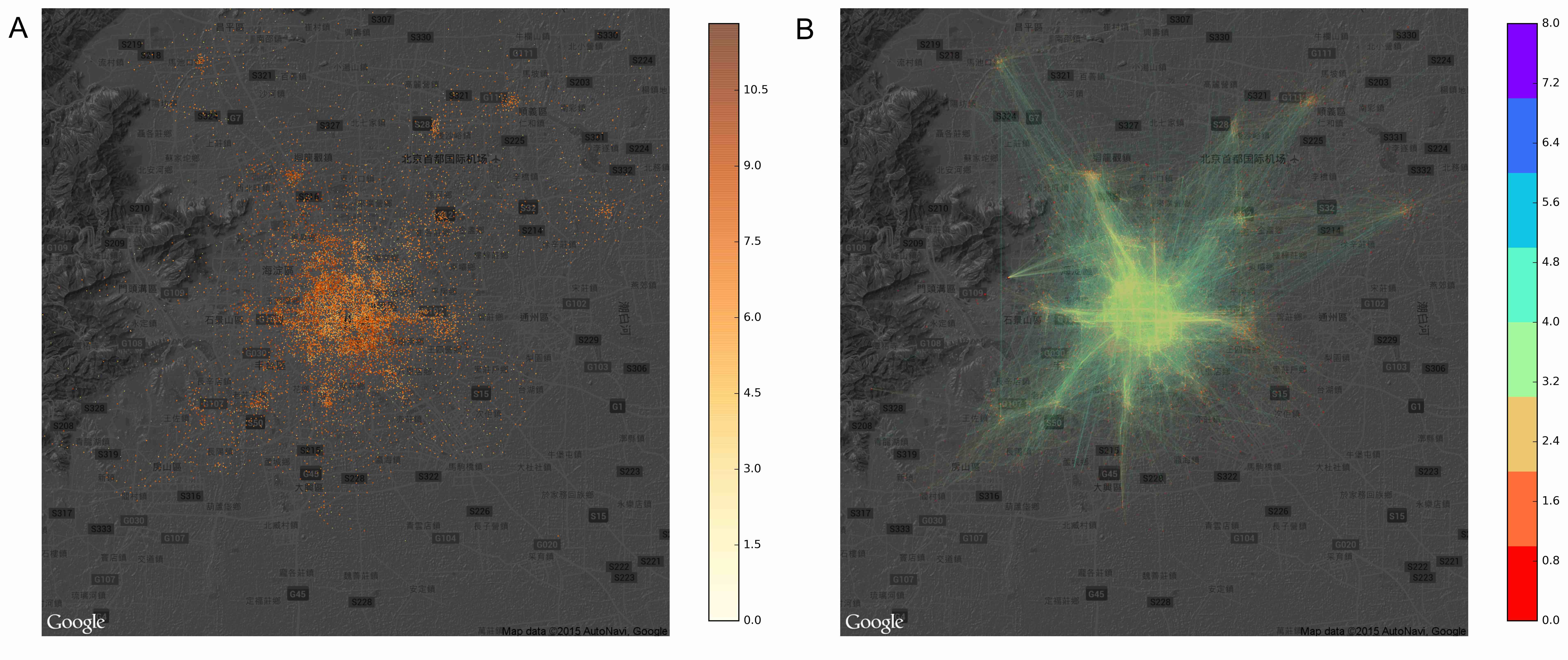}
    \caption{Spatial Distribution of Base Stations and Human Mobility. A. The distribution of 14,909 base stations. The darkness of the data points is proportional to the logarithm value of the mobility traffic of the base stations. B. The mobility of users in the physical space.}
    \label{spatial_mobility}
 \end{figure*}

\section*{Materials and Methods}
\subsection*{Data}
Mobile phone provides a new lens to investigate the human behaviors. Eagle et al. analyzed the relationship between mobile phone use and the social-economic development of British cities\cite{eagle2010network}. Song et al. used the mobile phone data to study the predicability of human mobility\cite{song2010limits}. In this study, we primarily work on a fine-grained anonymous smartphone dataset collected from a randomly sampled $1 \times 10^5$ users in a major city of China. The dataset profiles what websites or mobile apps users subsequently use in the virtual world as well as the physical locations. The dataset is anonymized, and we do not have the personal information of these users. 

Based on these information, we construct two networks to trace how users move from one website to another (attention network) and from one location to another (mobility network). In the mobility network, the nodes are mobile base stations ($N_m = 9899$), and the edges ($E_m = 39083$) are users travel from one base station to another. In the attention network, the nodes ($N _a= 16476$) are websites, and edges ($E_a = 144909$) represent the switch of users between websites. Figure \ref{spatial_mobility}A shows the spatial distribution of base stations. The base stations geographically distributed according to the shape of the city, and there are obvious clustering patterns. Figure \ref{spatial_mobility}B shows the mobility network with heatmaps.  The trajectory of human mobility vividly shows the transportation backbone of the city. 

 \begin{figure*}[!ht]
    \centering
    \includegraphics[scale=0.1]{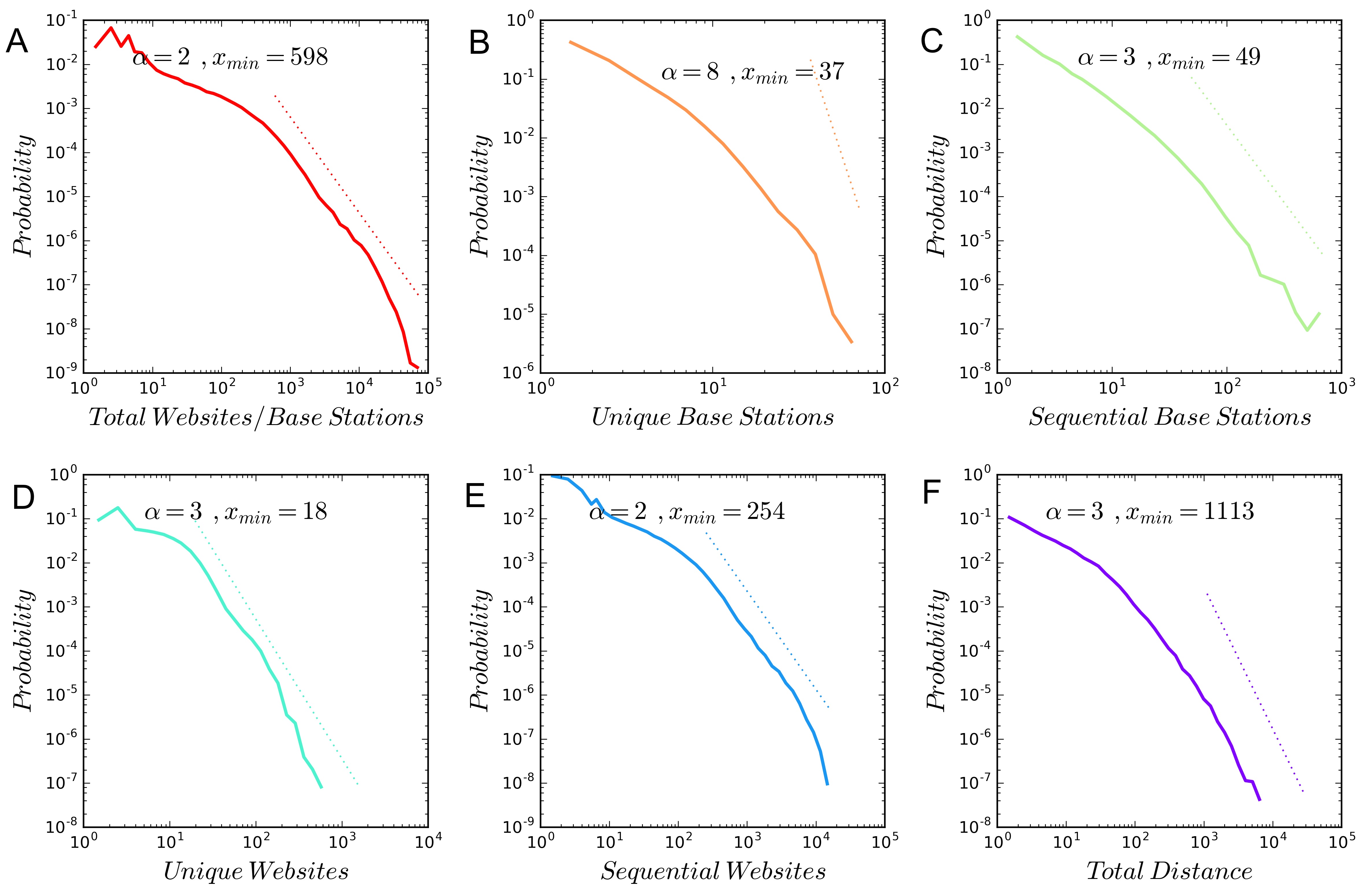}
    \caption{Long-tail Distribution of Human Behaviors}
    \label{power_law}
 \end{figure*}

To further describe the data, we calculate the following six variables for each individual based on their daily behavioral records: (1) the total number of records (websites/base stations); (2) the number of unique base stations; (3) the number of sequentially visited base stations (any two successive base stations are different); (4) the total distance of navigation; (5) the number of unique websites; (6) the number of sequentially visited websites. It turns out that the probability distributions of all these variables are featured by the long-tail. We fit their power law exponents using the powerlaw module \cite{alstott2014powerlaw}, and the results are shown in the Figure \ref{power_law}. Apparently, the human navigation in both the real world and the virtual space are very heterogeneous. 

\begin{figure*}[!ht]
    \centering
    \includegraphics[scale=0.07]{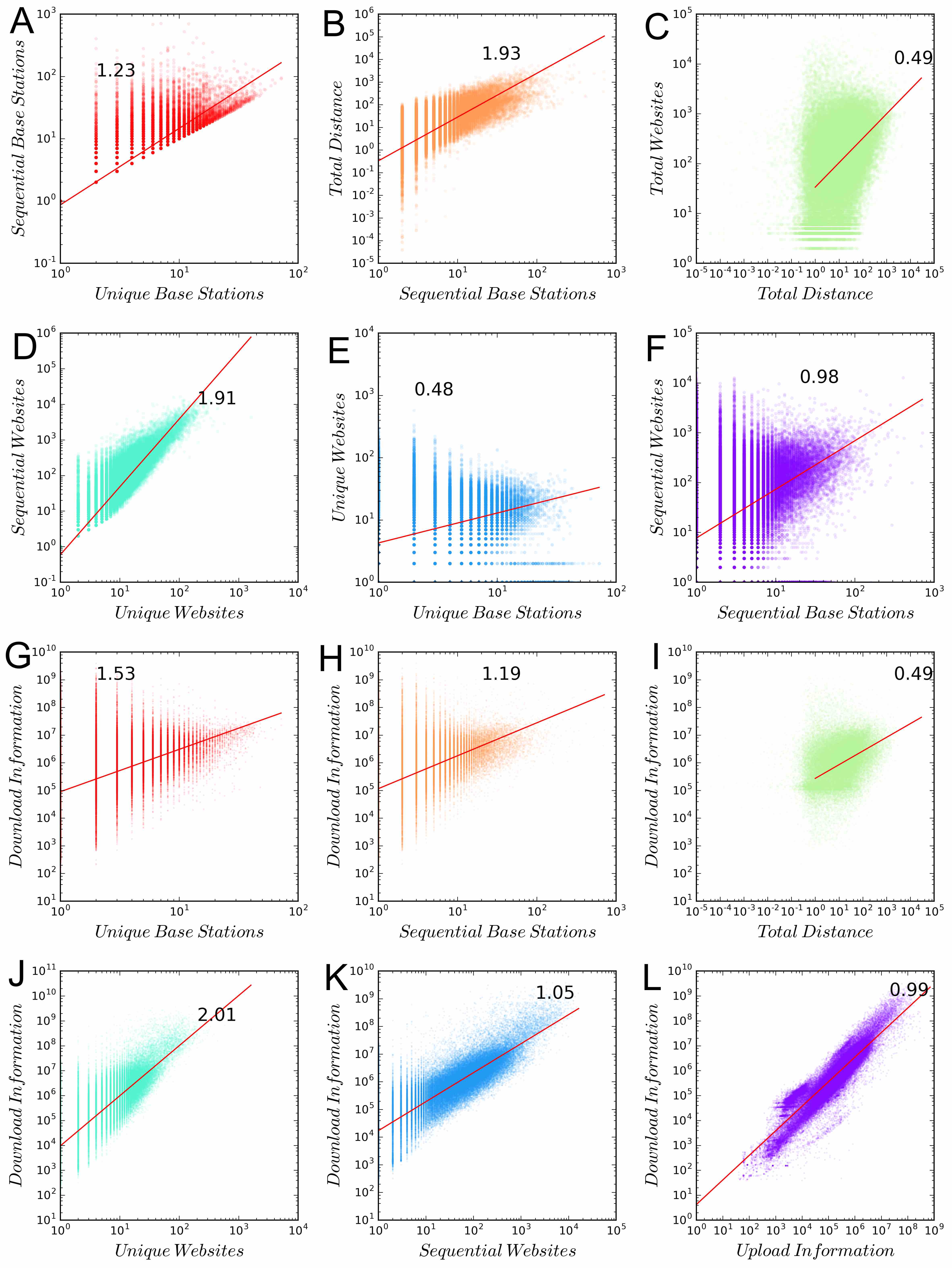}
    \caption{The Correlation Between the Virtual and Real World Navigation}
    \label{correlation_navigation}
 \end{figure*}

It is important to check how online behaviors correlate with the offline behaviors. We find that several pairs of variables are positively correlated (see Figure \ref{correlation_navigation}). In addition to that, we investigated the effect of several variables on the amount of information download within one day. The total distance of real world movement positively with the total number of websites visited within one day (Figure \ref{correlation_navigation}C), and the amount of information download(Figure \ref{correlation_navigation}I). The amount of information download positively correlates with the unique base stations (Figure \ref{correlation_navigation}G), sequential base stations (Figure \ref{correlation_navigation}H). The findings above reveals that those who are active online are largely proportionally active offline. Yet, it does not mean that online and offline human behaviors are similar. 

\subsection*{Network Renormalization}

We implement the box-counting method to renormalize the mobility network and the attention network \cite{song2005self}. The network is covered with boxes such that the nodes within a box are at a distance smaller than the length of box $l_B$. With the increasing of $l_B$, we need a smaller number of boxes $N(l_B)$ to cover the whole network. After we tile the entire network, each box was replaced by a single supernode, and connections between the supernodes are merged \cite{song2005self}. The renormalization procedure is repeated until there is only one node left in the network. 

\subsection*{TF-IDF}

We use the term frequency inverse document frequency (i.e., TF-IDF) measurement to find what the websites simultaneously visited by the moving citizens. TF-IDF is a statistic widely adopted to measure the importance of a term to a document in a corpus. It is the product of two statistics, term frequency and inverse document frequency. $TF$ measures the frequency of a term appearing in a given document. If a term appears many times in a document, its $TF$ value of this document will be large. $IDF$ measures the concentration of terms across documents. If that term appears in many documents, its $IDF$ value would be small. Combining the information supplied by both $TF$ and $IDF$, we can select the most important words for a given document. In our case, we can detect which websites are important for a group of users moving around a group of mobile base stations.

\section*{Acknowledgment}
The authors thank Zhang Chang for the assistance in providing data and inspiring discussion. 
We acknowledge financial support for this work from the National Social Science Foundation, grant number 15CXW017. 

\section*{Author contributions}
L.W and C.W designed research; C.W and L.W performed research; L.W and C.W collected and analyzed data; C.W and L.W wrote the manuscript; all authors discussed the results and reviewed the manuscript.

\section*{Competing financial interests}
The authors declare no competing financial interests.


\end{document}